\documentclass{appolb}
\usepackage{graphicx}
\begin{document}
\title{Exotics at Belle and perspectives at Belle II
\thanks{Presented at the Excited QCD Conference 2018, Kopaonik, Serbia}
}
\author{Elisabetta Prencipe, on behalf of the Belle and Belle II Collaborations
  \address{IKP1, Forschungszentrum J\"ulich, Leo Brandt strasse, 52428 J\"ulich}
}

\maketitle
\begin{abstract}
The search for multi-quark states beyond the constituent quark model (CQM) has resulted in the discovery of many new exotic states, starting with the observation of the X(3872), discovered by Belle in 2003. Also in the sector of charm-strange physics the CQM does not seem to describe properly all spectrum, despite of theoretical expectations. These new forms of quark bounds clearly show that mesons and baryons are not the only possibilities to be considered. We shortly report in this paper selected recent results on searching for such states at Belle, with the perspectives in the hadron physics program at the Belle II experiment.
\end{abstract}
  
\section{Introduction}
The Gell-Mann Zweig idea, known as the Costituent Quark Model~\cite{eli1, eli2} (CQM), classifies all known hadrons, and is still valid after more than half century, although in the meantime the quark $top$ and $bottom$ have been added to the pattern rising the number of quark families to 3. QCD- (Quantum Chromodynamics) motivated models based on this idea predict the existence of more complex structures than simple mesons (2 quark-bound states) or baryons (3 quark-bound states). Since the observation of the $X(3872) \rightarrow J/\psi \pi^+ \pi^-$~\cite{eli2bis}, and subsequent confirmations~\cite{eli3, eli4, eli5, eli6, eli7, eli8, eli8bis} it became evident that the Charmonium-like states with more than 3 quarks exist, and a flood of new theories and possible interpretations have been brought up.

Until 2003 theoretical models and experiments have got an overall agreement of 2-3 MeV/$c^2$ precision in the mass measurements of Charmonium states. After the observation of the X(3872) in 2003, from experimental point of view, lot of effort was put in searching for new forms of bound states. Some of them well fits the theory, while others are found at unexpected mass values, and still several possible interpretations are opened, mostly due to the poor available statistics in the past experiments to investigate them,  $i.e.$ performing  a full amplitude analysis, or detector limitations did not allow to measure the very narrow width, crucial to discriminate among different theoretical interpretations~\cite{elinora}. Undoubtedly, the  Belle experiment~\cite{elibelle} gave an important and substantial contribution to the field. Meson, baryons, dibaryons, glueballs, hybrids, tetraquarks, pentaquarks are only some of the hypoteses  populating the Charmonium-like literature.

\section{Z-charged States at Belle}
The Belle experiment ran from 1999 until 2010 at KEK, Tsukuba, Ibaraki prefecture (Japan). It was an asymmetric $e^+e^-$ experiment, collecting data at the center-of-mass (c.m.) energy of 10.56 GeV/$c^{\rm 2}$, $e.g.$ the threshold for the $\Upsilon(4S)$ production, decaying to $B \bar B$ meson pair. In addition to those data, it collected also samples at the c.m. energy of $\Upsilon(1,2,3,5)S$ production, resulting in 1.0 ab$^{-1}$ total integrated luminosity. The main goal of the Belle physics was the search for angles and sides of the Unitarity Triangle; but also rare decays, search for dark matter, and exotics in Charmonium and Bottomonium represent a substantial part of the Belle physics program.

In 2008 the first charged exotic state was announced by the Belle collaboration~\cite{eli10, eli10bis}, then confirmed by the LHCb collaboration~\cite{eli11}. Here we are definitively talking of an exotic state, as Charmonium states are supposed to be neutral by definition. In an update of the analysis performed in 2015~\cite{eli11bis}, the Belle collaboration published on 772 million $B \bar B$ pair the new mass and width values of the so-called Z(4430), $e.g.$ M = ($4485 \pm 22 ^{+28}_{-11}$) MeV/$c^2$ and $\Gamma$ = ($200^{+21}_{-46}$$^{+26}_{-35}$) MeV, observed with 6.4$\sigma$ significance, and provided the quantum number $\rm{J^P}$ = $\rm{1^+}$.

The observation of the Z(3900)~\cite{eli12} is also a milestone in the search for exotics at Belle. It followed the observaton of the same resonance at BES III~\cite{eli13}. The Belle analysis was performed using an integrated luminosity of 967 fb$^{-1}$, almost the full Belle data set. Belle observed the $Z(3900) \rightarrow \pi^- J/\psi$ with a significance larger than 5.2$\sigma$. Later on, in the invariant mass system of $\pi^- \psi$' a new exotic state was announced by Belle, the so-called Z(4050)~\cite{eli14}: in this case running over a sample of 980 fb$^{-1}$ integrated luminosity  Belle obtained the evidence of this state.

One can question what has been understood in 10 years of Z-exotic analyses. In trying to understand the pattern, we can attempt to classify those states in 2 main categories: those with large width, seen in B meson decays, apparently not connected to thresholds ($e.g.$ the Z(4430), the Z(4200)); and those directly seen in $e^+e^-$ interactions, characterized by narrow widths of the order of tens MeV ($e.g.$ the Z(3900), the Z(4020)). The first can be confirmed by LHCb, the latter from BES III. Belle and Belle II are in a unique position, as they can look for both Z-types in only one experiment.

\section{Observation of the X$^{*}$(3860)}
\begin{figure}[htb]
\centerline{%
\includegraphics[width=7cm]{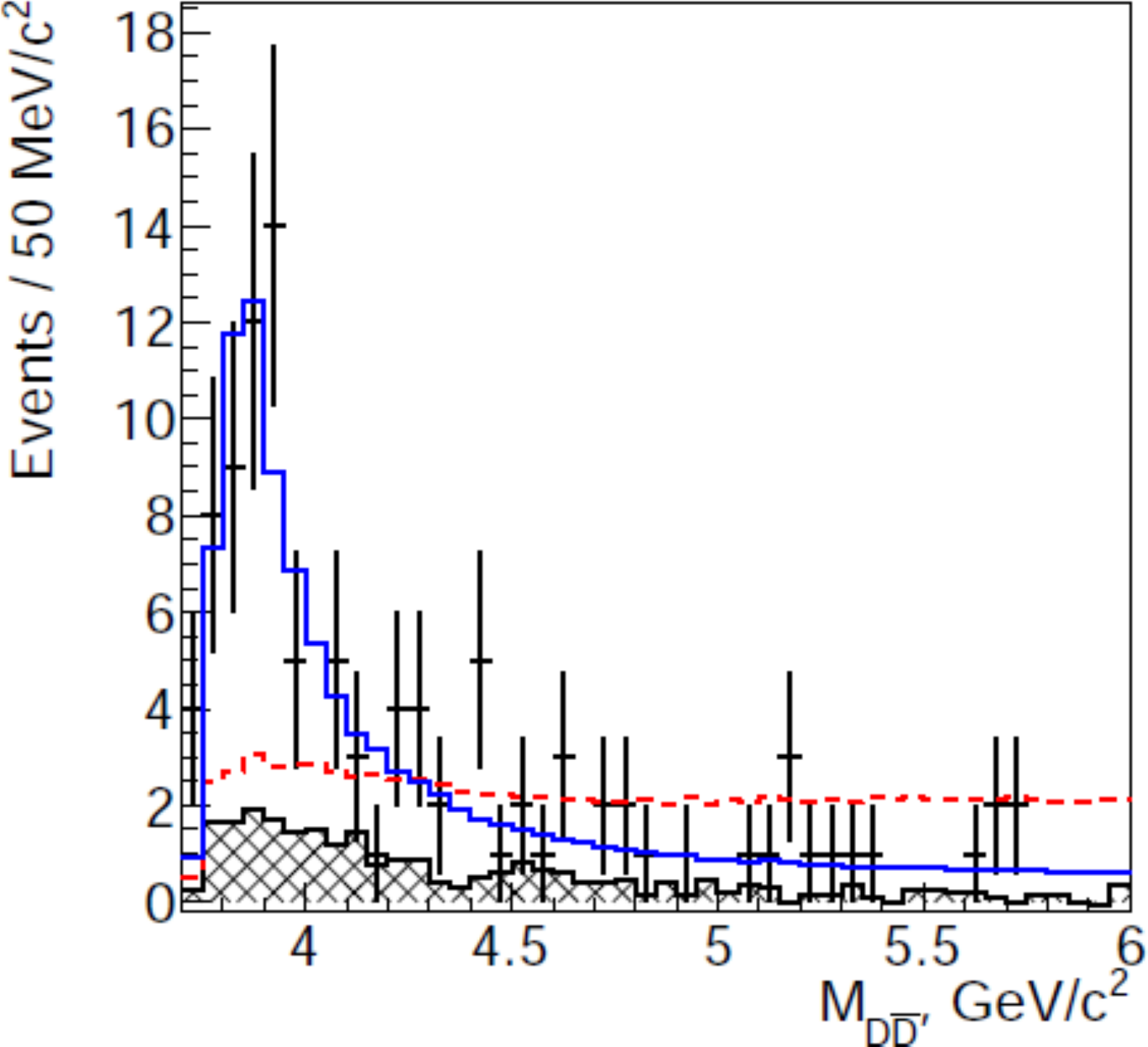}}
\caption{Projections of the signal fit results in the default model onto $M_{D \bar D}$.  The points with error bars are the data, the hatched histograms are the background, the blue solid line is the fit with a new $X^*$ resonance ($\rm{J^{PC}}$ = $0^{++}$) and the red dashed line is the fit with nonresonant amplitude only.}
\label{Fig:EliFig1}
\end{figure}

Recently Belle performed a new analysis: the investigation of the $D \bar D$ invariant mass in the $e^+e^- \rightarrow J/\psi D \bar D$ process~\cite{eli14}, using  980 fb$^{-1}$ integrated luminosity (see Fig.~\ref{Fig:EliFig1}). The partial wave analysis performed over so huge statistics allowed to establish a new resonance  at threshold, with $\rm{J^{PC} = 0^{++}}$ hypothesis favourite over 2$^{++}$, though the latter is not completely excluded. This new resonant state is a good candidate to be identified as the $\chi_{c0}(2P)$ in the Charmonium spectrum, assuming  the Z(3900) as the $\chi_{c2}(2P)$ and the X(3872) as the $\chi_{c1}(2P)$. The mass and the width are  measured, and equal to M = ($3862^{+26}_{-32}$$^{+40}_{-13}$) MeV/$c^2$ and $\Gamma$ =   ($201^{+156}_{-67}$$^{+83}_{-82}$) MeV.

\section{Search for Pentaquarks at Belle}
\begin{figure}[htb]
\centerline{%
\includegraphics[width=10cm]{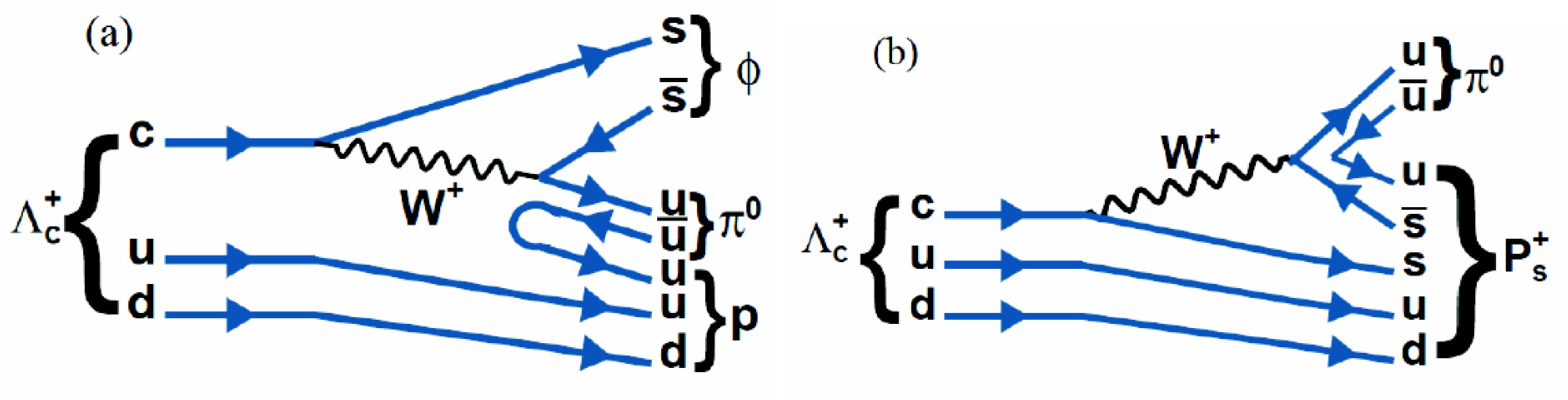}}
\caption{Feynman diagram for the decay (a) $\Lambda_c^+ \rightarrow \phi p \pi^0$ and (b) $\Lambda_c^+ \rightarrow P_s^+ \pi^0$.}
\label{Fig:EliFig2}
\end{figure}

The search for pentaquark states at Belle has been carried on, even if up to now did not give positive answer, but only upper limits. In a recent analysis performed over 915 fb$^{-1}$ integrated luminosity the Belle collaboration looked for the decay of $\Lambda_c^+ \rightarrow \phi p \pi^0$~\cite{eli15}, where a 5-quark state $P_s^+$ is supposed to decay to $\phi \pi^0$ (see Feynman diagram at Fig.~\ref{Fig:EliFig2}). It was measured a branching ratio (BR) upper limit at 90\% confidence level equal to $\cal B$($\Lambda_c \rightarrow P_s^+ \pi^0$)$\times \cal B$ ($P_s^+ \rightarrow \phi p$)$ <$8.3$\times$10$^{-5}$. This value is consistent with the BR estimated by LHCb in the analysis cited as Ref.~\cite{eli16}, although the decay analyzed by LHCb is different from that at Belle, since Belle cannot recontruct $\Lambda_b$, but only $\Lambda_c$ decays.

\section{From the experiment Belle to the experiment Belle II}
\begin{figure}[htb]
  \centerline{%
      \includegraphics[width=5cm]{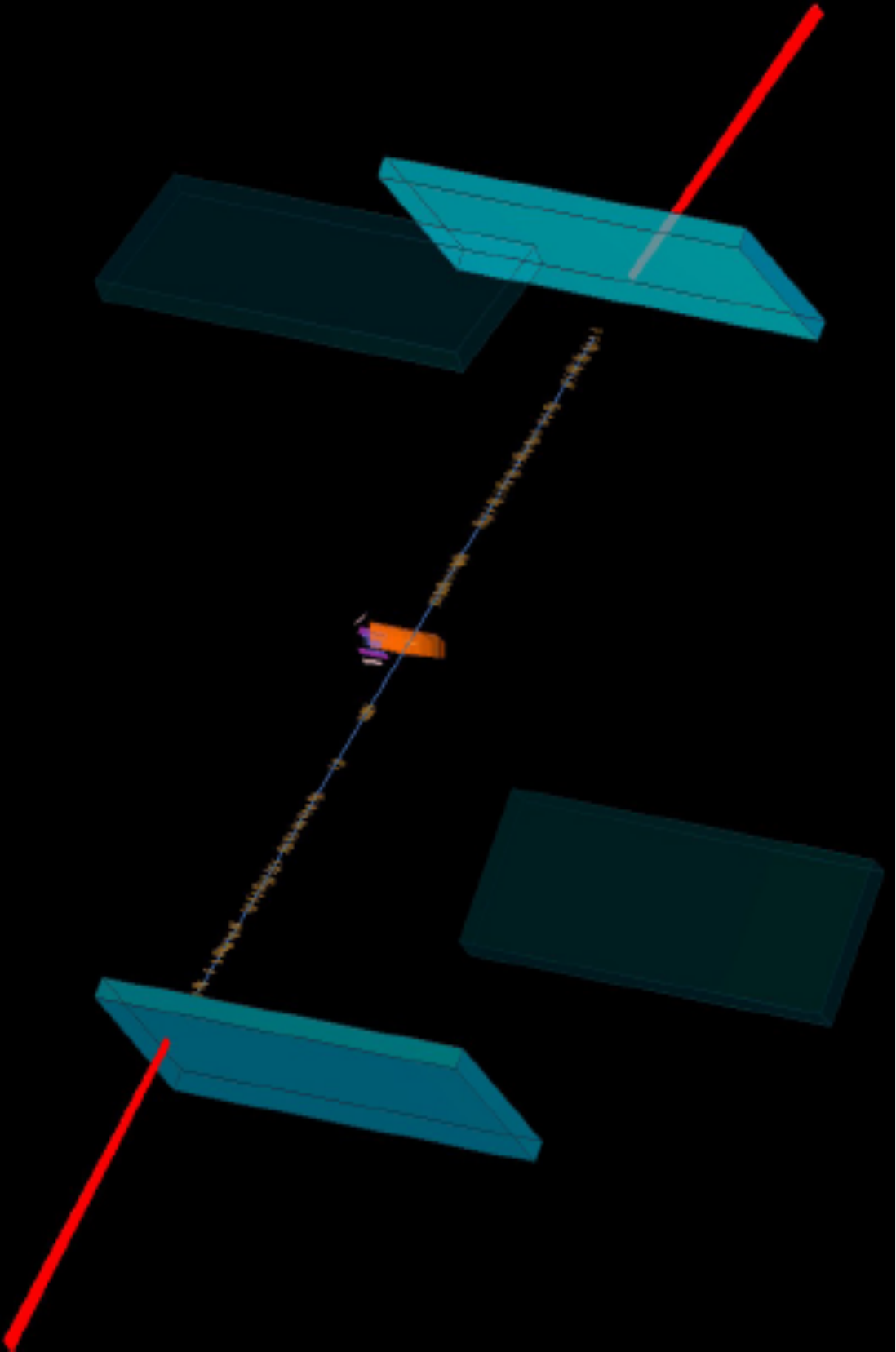}} 
    \caption{Cosmics in the PXD. Only 4 layers are mounted at the beginning of Phase II.}
\label{Fig:EliFig3}
\end{figure}

All recent analyses performed by Belle on exotic states have been performed over almost the full available data set, and show statistics limitation, not allowing a conclusion. It is then definitively needed an upgrade of the machine in order to have more statistics. The project Belle II~\cite{eli17}  has been in construction over the past decade, and on $\rm{18^{th}}$ of  February 2018 finally the first cosmics were collected (see Fig.~\ref{Fig:EliFig3}) and first studies with data started. On $\rm{26^{th}}$ of  April 2018 the first collisions happened, testifying the good shape of this new project and a new era for the $e^+e^-$ colliders. The experiment Belle II is located in Tsukuba (Japan), at the center of high energy physics KEK, and it is similar to Belle, with some substantial upgrade:

\begin{itemize}
\item a Pixel detector (PXD) has been planned, with vertex resolution in z-direction a factor 2 better than at Belle: from 50 $\mu$m (Belle) to 25 $\mu$m (Belle II);
\item the Time-of-Propagation (TOP) detector has been installed and properly working for particle identification purposes. The time resolution is 50 ps by design, and the detector surface is polished at nanometer precision;
\item KLM detector for $K_L$ and muon detection: 2 inner layers of barrel and all layers in the endcap replaced by scintillators;
\item the electromagnetic calorimeter (ECL) readout electronics has been exchanged: now fast ADCs are used;
\item a gain-factor 40 better than Belle in luminosity is planned, due to the new value of the beam current (which gives a factor 2) and the nano-beam principle, which gives an improvement of a  factor 40. In this way we expect 50 times more data than what was collected at Belle over 11 years, so in 2026 the recorded integrated luminosity is expected to be 50 ab$^{-1}$.  
\end{itemize}

With 1 ab$^{-1}$ planned in 2021, and 50 ab$^{-1}$ planned in 2026, Belle II is the most powerful $e^+e^-$ collider in the world, and great achievements are expected in spectroscopy (and not only).

\section{XYZ expectation at Belle II}
With the above expected high luminosity, Belle II can improve for sure some of the measurements already performed by Belle, and look for new still undisclosed forms of exotic matter. First of all, we mention the search for Y-vector states in ISR mechanism, for which Belle II is in a unique position. By extrapolating the yield of Y(4260) from the Belle publication as in Ref.~\cite{eli18}, in 2026 we expect to measure 29600 yield, for example. This opens the possibility to search for more rare Y(4260) decays, up to know not possibile due to the limited statistics. The search for the X(3872) radiative decays, $i.e.$ $X(3872) \rightarrow J/\psi \gamma$, is also a unique physics case at Belle II, since the reconsturction of low energy photons would not be a problem. The search for new Z-charged exotic states is also part of the Belle II program: with so high statistics amplitude analysis can be performed and the quantum numbers can be determined.

A unique physics case for Belle II would be the search for bottomonium states: $\Upsilon$(6S) running will be possible, and a program is already established to search for bottomonium in $\Upsilon$(5S) and $\Upsilon$(6S) transitions: for the first time radiative transitions between bottomonium states will be possible at Belle II, which search was limited at Belle due to low statistics.

\section{Comparison with other future facilities and Summary}
In summary, the Belle II experiment just started its data taking and it is in good shape. Original measurements are expected in the field of spectroscopy, especially for radiative decays, $\Upsilon$(5S, 6S) transitions, and ISR processes.

A comparison with other running and future experiments as conclusion here is provided, to help in understanding the performance and future opportunities of the Belle II experiment in spectroscopy. Table~\ref{tableEli1} compares the performance of Belle II, BES III, PANDA, and LHCb at one day of data taking. The reader should notice that the extrapolated yield for the X(3872) at Belle on the full data set is ~0.2, compared to 8.5 yield at Belle II reported in the table.

\begin{table}[!htb]
  \centering
  \caption{\label{tableEli1} Extrapolation of 1 day of data taking and comparison among different experiments, using the results in the  publications ~\cite{eli2bis, eli8bis, eli10, eli11, eli12, eli13, eli18, elihadron}. Information is extrapolated by assuming 40 fb$^{-1}$ integrated luminosity per day at Belle II,  1$\times$10$^{31}$ cm$^{-2}$ s$^{-1}$ luminosity in the startup mode of the future PANDA (the number in parenhtesis indicates the lower limit in the yield estimation, while the other number is the upper limit, evaluated as in Ref.~\cite{elihadron}), and assuming 2 fb$^{-1}$/year at LHCb. Where no number is quoted, the reader should assume no official paper can be cited for such extrapolation and/or it is not feasible. The Belle II yields are extrapolated from the corresponding  Belle analysis results.}
  \vspace{4 mm}
  \begin{tabular}{lrcccccl}\hline \hline 
   
    & Belle II   & BES III     &  LHCb   & PANDA\cr \hline \hline
    X(3872)       &  9  &      1 (radiative) &  2 (trigger) & 65\cr \hline
    Y(4260)       &  24 &  50                & -            & 1900 (67)\cr \hline
    Z(3900)       &   5 &   10               & -            & 405 (14)\cr \hline
    Z(4430)       &   8 & -                  & 5            &   -\cr \hline 
  \end{tabular} 
\end{table}

\end{document}